\documentclass[preprint,12pt]{elsarticle}

\usepackage{amssymb}
\usepackage{xcolor}
\usepackage{graphicx}
\usepackage{caption}
\usepackage{subcaption}
\usepackage{newtxtext}
\usepackage{newtxmath}
\usepackage{subcaption}

\journal{Journal of Fluids and Structures}

\begin{document}

\begin{frontmatter}



\title{Non-dimensional parameters for the flapping dynamics of multilayered plates}

\author[inst1]{Esha Jain}
\author[inst1]{Aditya Karthik Saravanakumar}
\author[inst2]{V. Joshi}
\author[inst1]{P. S. Gurugubelli}
\ead{pardhasg@bits-hyderabad.bits-pilani.ac.in}
\affiliation[inst1]{organization={Computing Lab, Department of Mechanical Engineering},
            addressline={Birla Institute of Technology and Science - Pilani, Hyderabad Campus}, 
            city={Secunderabad},
            postcode={500078}, 
            state={Telangana},
            country={India}}

\affiliation[inst2]{organization={Department of Mechanical Engineering},
            addressline={Birla Institute of Technology and Science - Pilani, K. K. Birla Goa Campus}, 
            city={Sancoale},
            postcode={403726}, 
            state={Goa},
            country={India}}

\begin{abstract}
The coupled flow-induced flapping dynamics of flexible plates in uniform axial flows are governed by three non-dimensional numbers, namely: Reynolds number $(Re)$, mass-ratio $(m^*)$, and non-dimensional flexural rigidity $(K_B)$. The traditional definition of these parameters is limited to isotropic single-layered flexible plates. However, there is a need to define these parameters for a more generic plate of multiple isotropic layers placed on top of each other. In this work, we derive the non-dimensional parameters for a flexible plate of $n$-isotropic layers and validate the non-dimensional parameters with numerical simulations. The proposed non-dimensional framework connects the experimental, numerical, and analytical studies on single-layered plates with the flapping dynamics of multilayered plates.
\end{abstract}


\begin{highlights}
\item Piezoelectric patches used for energy harvesting are usually multilayered.
\item Non-dimensional parameters for flapping instability of flexible multilayered plates.
\item Analytical derivation of equivalent flexural rigidity and equivalent mass ratio.
\end{highlights}


\end{frontmatter}


\section{Introduction}
\label{sec:sample1}
Fluid-structure interaction is a classical research area to investigate the effect of fluid flow on structures that can either move or deform and would in turn affect the fluid flow itself. Such interactions are omnipresent in nature (for example, the flapping of wings in avian animals and the fluttering of leaves) as well as engineering applications dealing with vortex-induced vibrations (VIV) \citep{Triantafyllou_2016_VIV, Gabbai2005} and bio-medicine \citep{Bluestein2008}. More recently, the problem of fluid-structure interactions has attracted substantial interest for its ability to harvest fluid kinetic energy into electrical energy. Electric energy is harvested from the vortex-induced vibrations (VIV) of circular cylinders as an ongoing research topic via piezoelectric \citep{zhang2019experimental, meliga2011extracting, zhang2017design, antoine2016optimal} and electromagnetic \citep{elvin2011experimentally, bernitsas2009vivace,soti2017harnessing,soti2018damping} effects. Apart from bluff bodies, the interaction of streamline bodies such as rigid foils and flexible plates \citep{zhang} with the incoming flow can be broadly categorized into propulsion \citep{quinn_2014,quinn_2014_1,heavingPropulsor,heaving_zhu_2013, Joshi_POF_2021} and energy extraction regimes \citep{allen2001,tang2009a,deniz2012, kinsey2006, Kinsey2012}.

Among the energy-harvesting alternatives, it has been noted that the maximum energy density of a piezoelectric transducer is around 63 and 8 times more than that of electrostatic and electromagnetic transducers, respectively \citep{JEON200516}. Furthermore, piezoelectric conversion has benefits such as high transmission efficiency and the least complex and economical design. An Eel-like device manufactured from polyvinylidene fluoride (PVDF) was utilized to harvest energy using ocean waves. It used a trail of traveling vortices behind a bluff body to strain piezoelectric elements for power generation \citep{allen2001, taylor_eel}. Similarly, a piezoelectric tree composed of a flexible and conducting leaf structure around a central ``trunk'' made of PVDF was used to harvest wind energy \citep{Li_Lipson_EnergyHarvesting}. Wind-induced vibrations lead to the deformation of the piezoelectric material, which results in the generation of electric current. Unlike rigid structures, modeling the flow across flexible structures is challenging numerically and experimentally. The strain energy of the flexible structure performing the flapping motion is converted into electrical energy via a piezoelectric mechanism in energy harvesting devices, which adds to the complexity of the modeling. Investigations have been conducted to optimize the energy output from such piezoelectric harvesting devices by studying the resonant conditions \citep{akaydin_2010}, flow orientation, plate orientation \citep{kim2013,gurugubelli_JFM,mittal_piezo_inverted,mittal_2016} and electrode position \citep{piezo_electrode_position}. In these experiments, it is common for the piezoelectric materials to be mounted on thicker plate-like structures made of materials such as Mylar \citep{akaydin_2010} or polyurethane \citep{allen2001}, thereby resulting in a multilayered flexible structure for energy harvesting. Various numerical methods based on the finite difference method \citep{peskin,connell2007}, finite element method \citep{tang_inverted_pof, gurugubelli_JFM} and lattice Boltzmann method \citep{lbm_flapping, lbm_parallel}, have been used to investigate the non-linear two/three-dimensional flow induced flapping response, force and vortex dynamics to understand the physics behind the flapping phenomena for isotropic flexible structures.

Even though piezoelectric energy harvesting devices are multilayered, the current understanding of the non-linear flapping dynamics is limited to results from isotropic models. An attempt was made earlier by \citet{deniz2012}, where the flapping dynamics of a three-layered flexible plate were numerically investigated by considering an equivalent single-layered plate's Young's modulus for the three-layered flexible plate. More recently, in \citet{multilayered_numerical}, a quasi-monolithic formulation was considered to numerically investigate the post-critical flapping dynamics of a two-layered plate in a uniform flow. The study revealed that variation in material properties across the layers could result in complex flapping dynamics. Hence, there is a crucial need for developing non-dimensional parameters that can generalize the flapping dynamics of multilayered structures.

Building upon the work of \citet{multilayered_numerical}, the current study aims at formulating the equivalent, non-dimensional mass ratio and flexural rigidity for an n-layered plate. Therefore, it lays the groundwork for a complete description of the flapping dynamics for multilayered plates. The following formulation simplifies the analysis of a system by reducing the number of variables and parameters involved. According to the knowledge of the authors, although multilayered plates are extensively used in engineering, there does not exist a non-dimensional framework to describe the physics of such sandwiched structures undergoing flapping due to interaction with the surrounding flow. Therefore, this generalized novel description of multilayered plates will be  imperative for their experimental, numerical and analytical analyses.  

The layout of the article is as follows. The non-dimensional parameters are derived and constructed for the multilayered plates in Section \ref{non_dimensional_parameters}. The generalized description is then systematically verified by considering multiple case scenarios in Section \ref{verification}. The study is summarized and concluded in Section \ref{conclusion}.

\section{Construction of Non-dimensional Parameters}
\label{non_dimensional_parameters}
\begin{figure}
        \centering
		\begin{subfigure}{0.59\textwidth}
			\centering
			\vspace{1.5cm}
			\includegraphics[width=0.99\columnwidth]{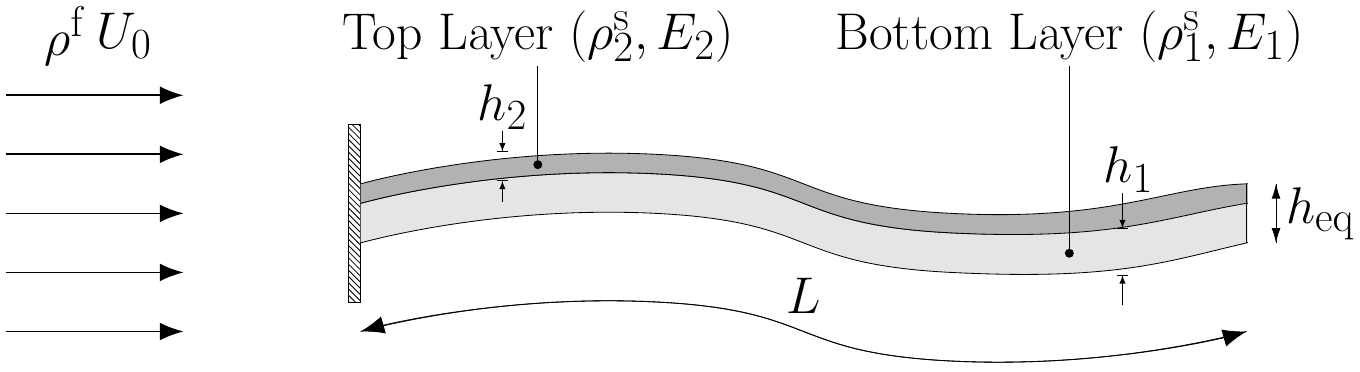}
					\vspace{1.6cm}
					\caption{Side view}
		\end{subfigure}	
		\begin{subfigure}{0.33\textwidth}
			\centering
			\includegraphics[width=0.99\columnwidth]{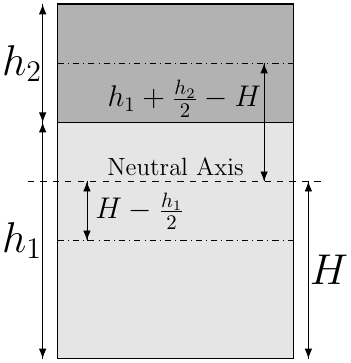}
			\caption{Cross-sectional view}
		\end{subfigure}	
		\caption{A schematic of a two-layered flexible plate, with a fixed leading edge, flapping in a uniform stream.}
		\label{Problem_Statement} 	
\end{figure}\noindent
In this section, we present an analytical derivation for the non-dimensional parameters that govern the self-sustained flapping dynamics of multilayered plates. Initially, we construct the non-dimensional parameters for a two-layered plate and then generalize the derivation for the case of an $ n $-layered plate. The flapping dynamics of a single-layered isotropic flexible plate are relatively well understood and are known to depend on the structure to fluid mass-ratio ($m^*$), non-dimensional flexural rigidity ($K_B$) and Reynolds number ($Re$) \citep{shelley05,connell2007,jaiman2013} which are defined as
\begin{equation}
	m^*=\frac{\rho^\mathrm{s}h}{\rho^\mathrm{f}L} \qquad K_B=\frac{B}{\rho^\mathrm{f}U_0^2L^3} \qquad Re = \frac{\rho^\mathrm{f}U_0L}{\mu^\mathrm{f}}, \label{numbers1}
\end{equation}
where $B$ and $\rho^\mathrm{s}$ represent the flexural rigidity and density of the flexible plate, respectively, and $\mu^\mathrm{f}$ is the dynamic viscosity of the fluid. These definitions can be extended for multilayered flexible plates by modifying the non-dimensional parameters,  $m^*$, and $K_B$, to replace the properties of the single-layered plate such as $\rho^\mathrm{s}$, $h$ and $B$ with their equivalent values for a multilayered plate.

Figure~\ref{Problem_Statement} shows a representative schematic of a two-layered flexible plate of length $L$ and thickness $h_\mathrm{eq}$ flapping about its leading edge. The flexible plate interacts with a uniform incompressible viscous fluid of density $ \rho^\mathrm{f} $, flowing along its length at a flow velocity of $U_0$. In this work, $ \left(\rho^\mathrm{s}_1, E_1, \nu_{1},h_1\right) $ and $ \left(\rho^\mathrm{s}_2, E_2, \nu_{2},h_2\right) $ represent the density, Young's modulus, Poisson's ratio and thickness of the bottom and top layers, respectively. The total thickness of the two-layered plate is given as $h_\mathrm{eq}=h_1+h_2$.

The structural dynamics of the flapping plate are described by the Euler-Bernoulli beam equation given as    
\begin{equation}
	\rho_\mathrm{eq}^\mathrm{s}h_\mathrm{eq}\frac{\partial ^{2}w}{\partial t^{2}}+B_\mathrm{eq}\frac{\partial^{4}w}{\partial x^{4}}=f,\label{eulerbernoulli}
\end{equation}  
\noindent where $\rho_\mathrm{eq}^\mathrm{s}$ and $B_\mathrm{eq}$ represent the equivalent density and equivalent flexural rigidity for the combined two-layered plate. Here, $w$ denotes the transverse deflection of the plate due to the fluid loading $f$ acting on the plate. The equivalent density $\rho_\mathrm{eq}^\mathrm{s}$ and equivalent flexural rigidity $B_\mathrm{eq}$ are defined as
\begin{gather}
	\rho_\mathrm{eq}^\mathrm{s}=\frac{\rho^\mathrm{s}_{1}h_{1}+\rho^\mathrm{s}_{2}h_{2}}{h_\mathrm{eq}} \quad
	B_\mathrm{eq}=\frac{E_{1}I_{1}}{(1-\nu_{1}^{2})}+\frac{E_{2}I_{2}}{(1-\nu_{2}^{2})}.\label{flexuralrigidity}
\end{gather}
In Eq.~\ref{flexuralrigidity}, $I_1$ and $I_2$ represent the area moment of inertia for the bottom and top layers, respectively, calculated with respect to the neutral axis of the combined two-layered plate. For an isotropic plate with a uniform cross-section, the neutral axis will always pass through the centroid of the plate's cross-section. However, for a two-layered or multilayered plate, the neutral axis may not pass through the centroid of the cross-section as its position depends on the thickness and Young's modulus of each layer. In such cases, the position of the neutral axis can be determined by considering the stress equilibrium condition in the streamwise direction at any cross-section. For the case of pure bending in a two-layered plate, the equilibrium condition can be formulated as
\begin{gather}
	\Sigma F_\mathrm{x}=\int _{A}\sigma _\mathrm{x}dA=0\\
	-\int _{0}^{h_1}E_{1}\frac{(y-H)}{\gamma}~{b}~dy-\int _{h_1}^{h_1 + h_2}E_{2}\frac{(y-H)}{\gamma}~{b}~dy=0,\label{force}
\end{gather}
\noindent where  $\sigma _\mathrm{x}$, $\gamma$, ${b}$ and $H$ represent the normal stress along the streamwise direction, the plate width, the radius of curvature and the distance of the neutral axis from the bottom of the plate, respectively. Eq.~\ref{force} can be solved to deduce the expression for the position of a neutral axis $H$ as
\begin{equation}
	H=\frac{E_{1}h_{1}\left({h_1}/{2}\right)+E_{2}h_{2}\left(h_1+{h_2}/{2}\right)}{(E_{1}h_{1}+E_{2}h_{2})}.\label{neutralsurface}
\end{equation}
The parallel axis theorem can then be used to calculate the area moments of inertia $I_{1}$ and $I_{2}$ about the neutral axis (see Fig.~\ref{Problem_Statement}). The resulting expressions are
\begin{equation}
	I_1=\frac{h_{1}^{3}}{12}+h_{1}\left (\frac{h_{1}}{2} -H \right )^{2} \hspace{0.5cm} I_2=\frac{h_{2}^{3}}{12} +h_{2}\left (h_1 + \frac{h_{2}}{2}-H \right )^{2}.\label{area_moment}
\end{equation}
\noindent Upon substituting these expressions in Eq.~\ref{flexuralrigidity}, we obtain the final form of the equation for the equivalent flexural rigidity ($B_\mathrm{eq}$) as 
\begin{equation}
	\begin{split}
	B_\mathrm{eq}&=\frac{E_{1}}{(1-\nu_{1}^{2})}\left [ \frac{h_{1}^{3}}{12}+h_{1}\left (\frac{h_{1}}{2} -H \right )^{2}\right ]+\frac{E_{2}}{(1-\nu_{2}^{2})}\left [ \frac{h_{2}^{3}}{12} +h_{2}\left (h_1 + \frac{h_{2}}{2}-H \right )^{2} \right ]\label{flexuralrigidity1}
	\end{split}
\end{equation}
Finally, the modified mass-ratio $m^*_{\mathrm{eq}}$ and non-dimensional flexural rigidity $K_B^{\mathrm{eq}}$ that govern the flapping dynamics of a two-layered flexible plate are defined as
\begin{equation}
	m^*_{\mathrm{eq}} = \frac{\rho^\mathrm{s}_\mathrm{eq}h_\mathrm{eq}}{\rho^\mathrm{f}L}\qquad K_B^{\mathrm{eq}} = \frac{B_\mathrm{eq}}{\rho^\mathrm{f}U_0^2L^3}.
	\label{modenumbers}
\end{equation}   
Furthermore, the expressions for $H$ and $B_\mathrm{eq}$ in Eqs.~\ref{neutralsurface}~and~\ref{flexuralrigidity1}, respectively, for the case of a two-layered plate can be generalized for a $n$-layered plate following the same equilibrium condition approach as
\begin{gather}
	H=\frac{E_{1}h_{1}\left({h_1}/{2}\right)+\cdots + E_{n}h_{n}\left(h_1+\cdots+h_{n-1}+h_n/2\right)}{(E_{1}h_{1}+E_{2}h_{2}+\cdots+E_{n-1}h_{n-1}+E_nh_n)} \notag \\[5pt]
	H=\frac{\sum_{i=1}^{n}\left [ E_{i}h_{i}\left (\sum_{j=1}^{i}h_{j} - \frac{h_{i}}{2} \right )\right ]}{\sum_{i=1}^{n} E_{i}h_{i}}
	\label{neutralsurface_nlay}\
\end{gather}
\begin{eqnarray}
	B_\mathrm{eq} & = & \frac{E_{1}}{(1-\nu_{1}^{2})}\left[ \frac{h_{1}^{3}}{12}+h_{1}\left (\frac{h_{1}}{2} -H \right )^{2}\right ]+\frac{E_{2}}{(1-\nu_{2}^{2})}\left [ \frac{h_{2}^{3}}{12} +h_{2}\left (h_1 + \frac{h_{2}}{2}-H \right )^{2} \right ]+ \cdots \notag \\
	 & + & \frac{E_{n}}{(1-\nu_{n}^{2})}\left[ \frac{h_{n}^{3}}{12}+h_{n}\left (h_1+\cdots+h_{n-1}+\frac{h_{n}}{2} -H \right )^{2}\right]
\end{eqnarray}
\begin{equation}
	B_\mathrm{eq}=\sum_{i=1}^{n}\frac{E_{i}}{1-\nu^{2}_{i}}\left [ \frac{h^{3}_{i}}{12} + h_{i}\left (\sum_{j=1}^{i}h_{j} - \frac{h_{i}}{2}-H \right )^{2}\right ] \label{flexuralrigidity1_nlay}
\end{equation}
We can use these definitions to define the non-dimensional parameters for a $ n $-layered plate, according to Eq.~\ref{modenumbers}. It is noteworthy that although the equivalent mass ratio is the linear summation of the independent mass ratios of each layer, the equivalent non-dimensional flexural rigidity is not equal to the sum of the independent non-dimensional flexural rigidity of each layer.

\section{Numerical Verification}
\label{verification}
The newly proposed non-dimensional parameters were validated by conducting ten parametric numerical experiments to simulate the flapping dynamics of a two-layered plate placed in a uniform stream for $Re=1000$. Table~\ref{governingpara} provides the material properties of the ten cases that can be further divided into two sets. Set-I consists of the first five cases from Table~\ref{governingpara} which will validate the equivalent flexural rigidity $ K_B^\mathrm{eq}$. Here the values of $m_{\mathrm{1}}^{*}$, $m_{\mathrm{2}}^{*}$ and $m_{\mathrm{eq}}^{*}$ are kept constant and equal to 0.1. For cases 1 and 2, the parameters $K_{\mathrm{B}}^{1}$ and $K_{\mathrm{B}}^{2}$ are selected such that $K_{\mathrm{B}}^{eq}$ for both cases is equal to 0.0005. Similarly, different values of $K_{\mathrm{B}}^{1}$ and $K_{\mathrm{B}}^{2}$ are chosen for cases 3 and 4 such that the resultant $K_{\mathrm{B}}^{eq}$ is 0.0004. In order to validate the effect of the thickness of the layers on the non-dimensional parameters, cases 1 to 4 are constructed by assuming the thickness of the top layer is the same as that of the bottom layer i.e. $h_1 = h_2 = 0.005L$  and case 5 is constructed by choosing non-equal thicknesses for the individual layers. Set-II comprises cases 6 to 10 that validate the equivalent mass ratio $m^*_\mathrm{eq}$. The values of $K_{\mathrm{B}}^{1}$, $K_{\mathrm{B}}^{2}$ and $K_{\mathrm{B}}^{eq}$ are kept constant and equal to 0.0001 for the full set. For cases 6 and 7, the parameters $m_\mathrm{1}$ and $m_\mathrm{2}$ are selected such that $m^*_\mathrm{eq}$ is equal to 0.075 for both cases. Similarly, different values of $m_\mathrm{1}$ and $m_\mathrm{2}$ are chosen for cases 8, 9, and 10 such that the resultant $m^*_\mathrm{eq}$ remains 0.05 for all three cases. Similar to Set-I, cases 6 to 9 assume that the thickness of the top and the bottom layers remain identical and equal to $0.005L$ and case 10 presents the condition of layers with different thicknesses.

\begin{table}
\caption{\label{} Summary of governing parameters considered for the numerical validation.}
\begin{tabular}{c|c|c|c|c|c|c|c|c}
				Case & $\frac{\rho_{1}^{\mathrm{s}}}{\rho^{\mathrm{f}}}$ & $\frac{\rho_{2}^{\mathrm{s}}}{\rho^{\mathrm{f}}}$ & $m_{\mathrm{eq}}^{*}$ & $h_1$ & $h_2$ & $\frac{E_{1}}{\rho^{\mathrm{f}}U_{0}^{2}\left(1-\nu_{1}^{2}\right)}$ & $\frac{E_{2}}{\rho^{\mathrm{f}}U_{0}^{2}\left(1-\nu_{2}^{2}\right)}$ & $K_{B}^{\mathrm{eq}}$\tabularnewline
                \hline 
                1 & 10.0 & 10.0 & 0.01 & 0.005 & 0.005 & 27761.5 & 1641.8 & 0.0005 \tabularnewline
                2 & 10.0 & 10.0 & 0.01 & 0.005 & 0.005 & 6000.0 & 6000.0 & 0.0005 \tabularnewline
                3 & 10.0 & 10.0 & 0.01 & 0.005 & 0.005 & 4800.0 & 4800.0 & 0.0004 \tabularnewline
                4 & 10.0 & 10.0 & 0.01 & 0.005 & 0.005 & 28560.4 & 775.8 & 0.0004 \tabularnewline
                5 & 10.0 & 10.0 & 0.01 & 0.0025 & 0.0075 & 778.5 & 10000.0 & 0.0004 \tabularnewline
                6 & 10.0 & 5.0 & 0.075 & 0.005 & 0.005 & 1200.0 & 1200.0 & 0.0001 \tabularnewline
                7 & 12.5 & 2.5 & 0.075 & 0.005 & 0.005 & 1200.0 & 1200.0 & 0.0001 \tabularnewline
                8 & 5.0 & 5.0 & 0.05 & 0.005 & 0.005 & 1200.0 & 1200.0 & 0.0001 \tabularnewline
                9 & 7.5 & 2.5 & 0.05 & 0.005 & 0.005 & 1200.0 & 1200.0 & 0.0001 \tabularnewline
                10 & 12.5 & 2.5 & 0.05 & 0.0025 & 0.0075 & 1200.0 & 1200.0 & 0.0001 \tabularnewline
			\end{tabular}
            \label{governingpara}
\end{table}

\subsection{Computational setup}
\begin{figure}
    \centering
    \includegraphics[trim={0mm 0mm 0mm 4mm},clip,width=0.99\columnwidth]{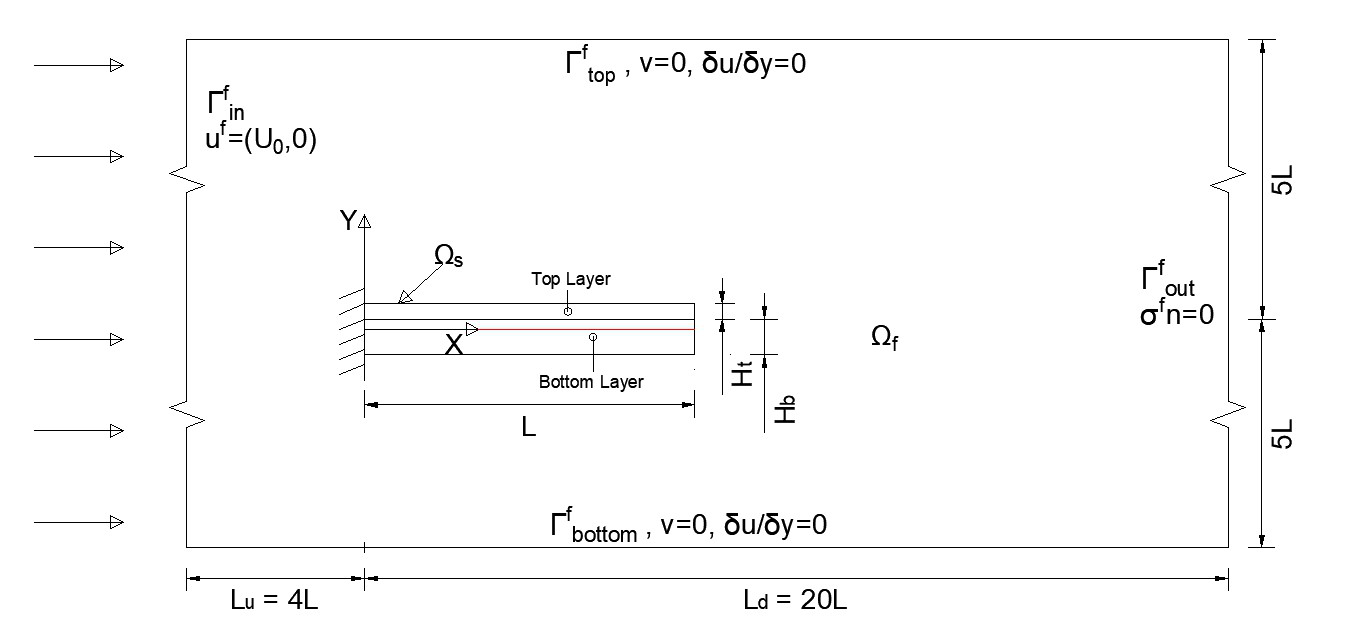}
		\caption{In the computational setup presented here, we can see the two-layered flexible plate in a uniform flow of velocity $U_0$, the boundary conditions and the interface separating the fluid domain from the solid domain. The dashed line (---) indicates the neutral surface of the plate.}
		\label{compset} 	
\end{figure}

\begin{figure}	
    \begin{minipage}[b]{0.5\linewidth}
		\includegraphics[width=0.99\columnwidth]{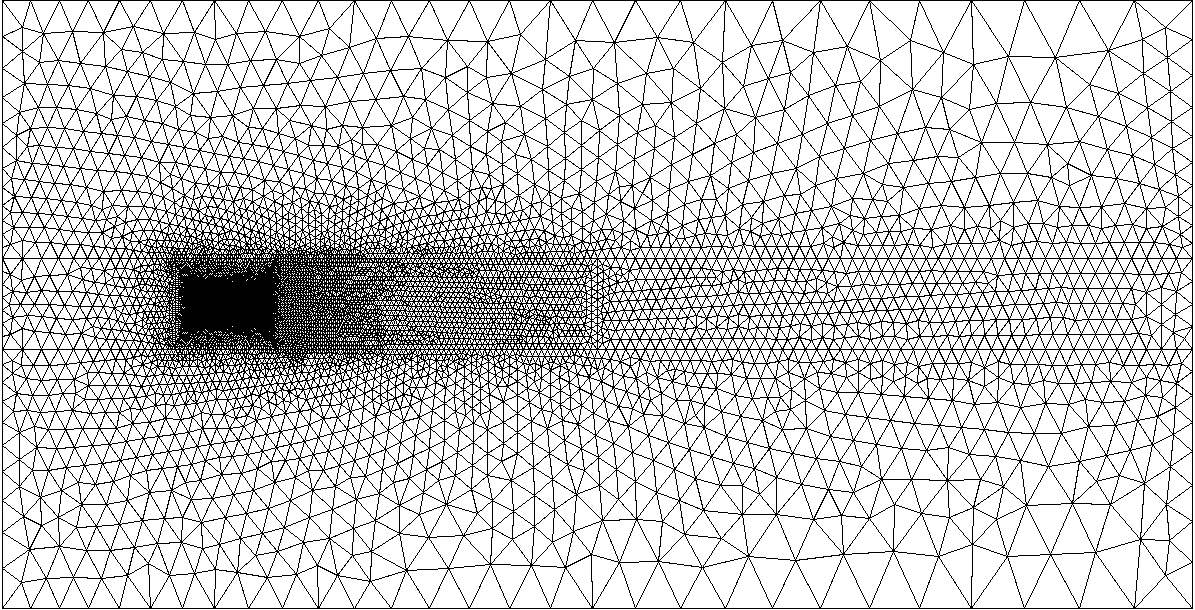}
    \end{minipage}	
    \begin{minipage}[b]{0.5\linewidth}
		\includegraphics[width=0.92\columnwidth]{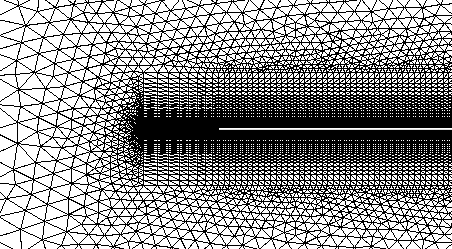}
    \end{minipage}
            \caption{The initial state of the finite element mesh in the fluid domain comprising 29549 nodes and 14657 elements (left) and the close-up view of the mesh at the leading edge (right)}
		\label{mesh} 	
\end{figure}

The interactions of the plate with the fluid flow are simulated using an in-house finite element based solver with exact interaction tracking presented in \citep{multilayered_numerical}. The computational setup shown in Fig.~\ref{compset} consists of a two-layered flexible plate clamped at the leading edge. The trailing edge is left free to undergo self-induced flapping due to interaction with a uniform axial flow. The computational domain considered for the study is of the size $24L\times10L$, wherein $L$ is the length of the plate. Considering that the fluid flows from left to right, a free stream velocity of $U_0$ enters the computation domain through $\Gamma^{f}_{in}$ and leaves through $\Gamma^f_{out}$. Accordingly, a traction-free boundary condition is placed on the outlet ($\Gamma^f_{out}$) of the computational domain. The top ($\Gamma^f_{top}$) and bottom ($\Gamma^f_{bottom}$) limits of the computational domain are bounded by the free-slip boundary condition. A no-slip boundary condition is imposed on the plate-fluid interface. 
The solver is $2^\mathrm{nd}$ order accurate in time and $3^\mathrm{rd}$ order accurate in space.  As per the mesh convergence study carried out earlier \citep{multilayered_numerical}, an identical high-order $\mathbb{P}_2$ finite element computational mesh has been selected. The mesh shown in Fig.~\ref{mesh} consists of 29549 nodes and 14657 $\mathbb{P}_2$ elements.

\subsection{Validation of equivalent flexural rigidity}
\begin{figure}
            \centering
            \begin{minipage}[b]{0.49\linewidth}
			\centering
			\includegraphics[trim={1.5mm 2mm 12mm 0mm},clip,width=0.99\columnwidth]{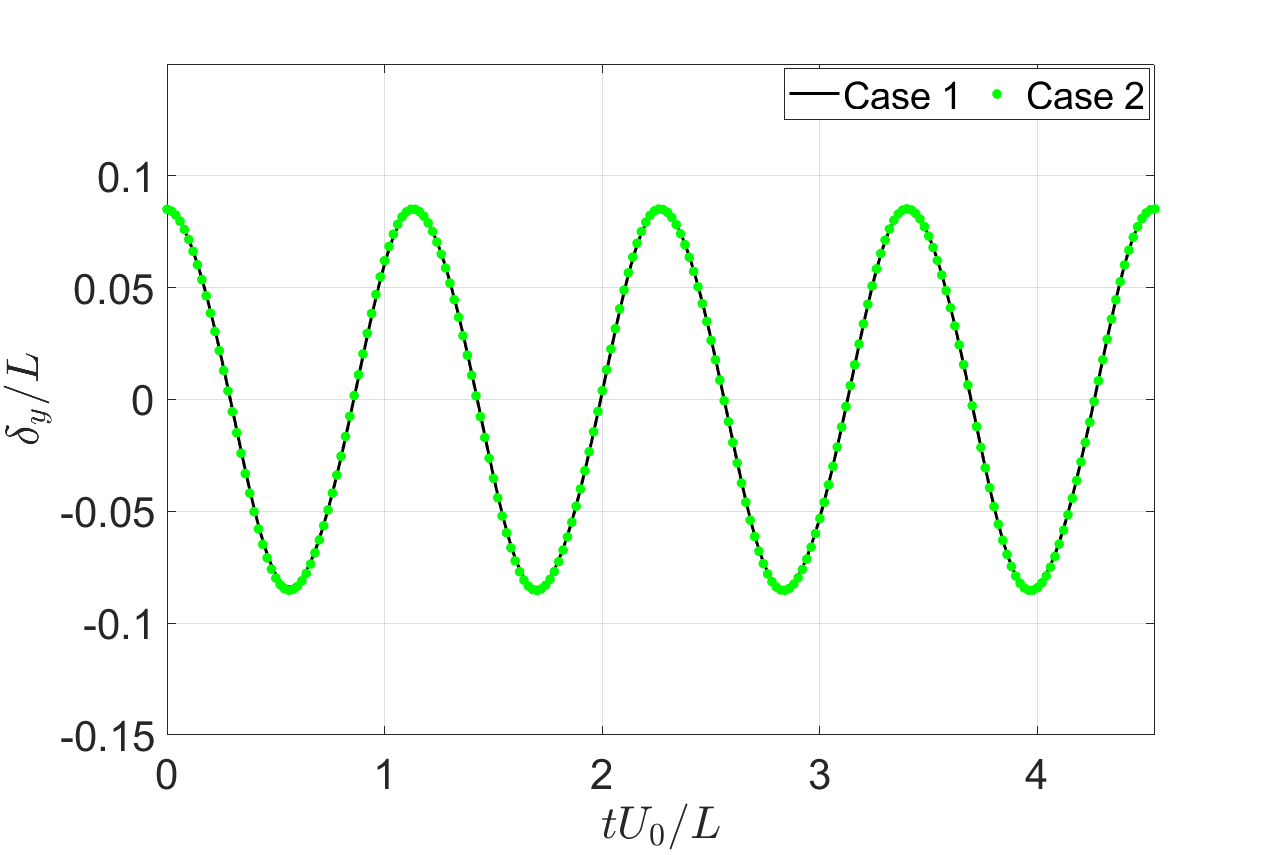}
		\end{minipage}
		\caption{Tip displacement time history for cases 1 and 2, plotted against non-dimensional time.}
		\label{tipdisp12} 	
\end{figure}

\begin{figure}		
            \begin{minipage}[b]{0.49\linewidth}
			\centering
			\includegraphics[trim={1.5mm 2mm 12mm 0mm},clip,width=0.99\columnwidth]{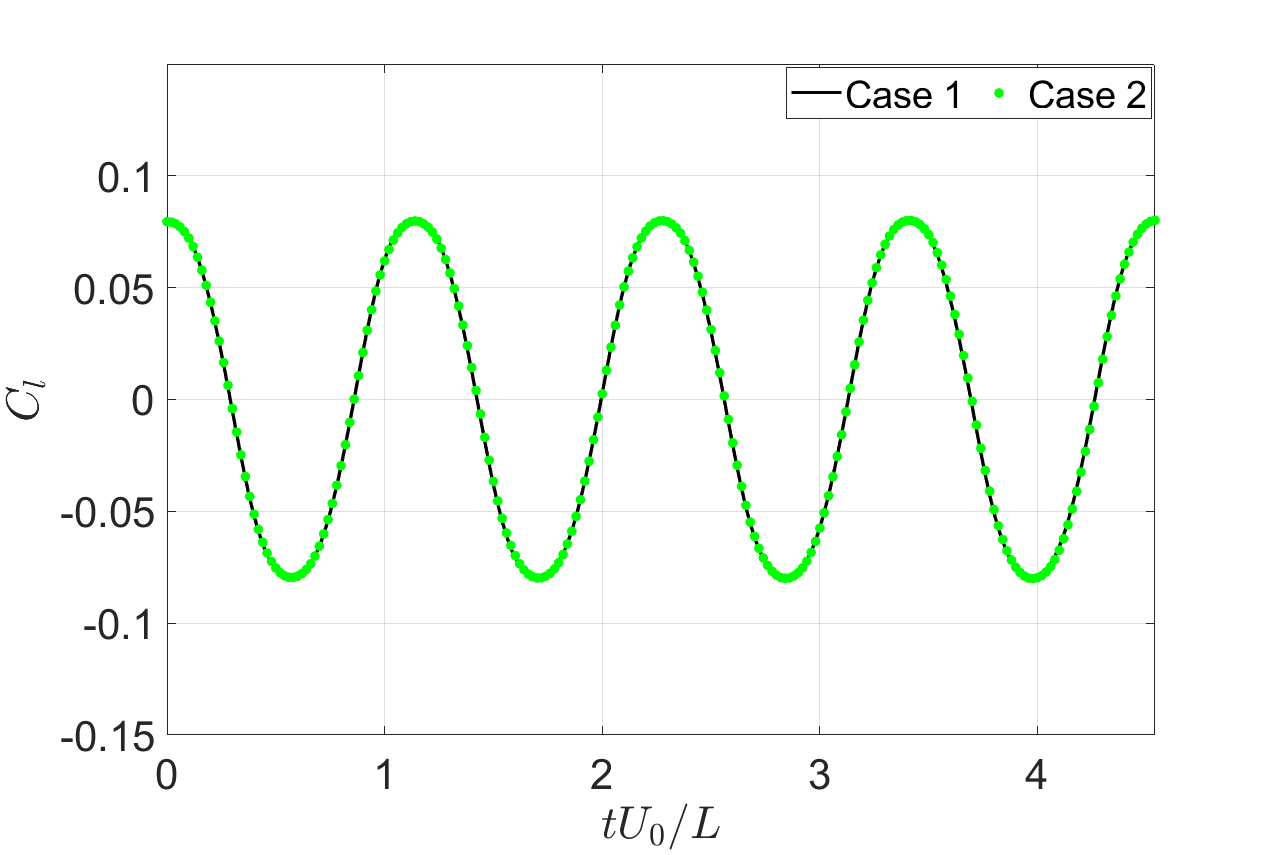}
		\end{minipage}
            \begin{minipage}[b]{0.49\linewidth}
			\centering
			\includegraphics[trim={1.5mm 2mm 12mm 0mm},clip,width=0.99\columnwidth]{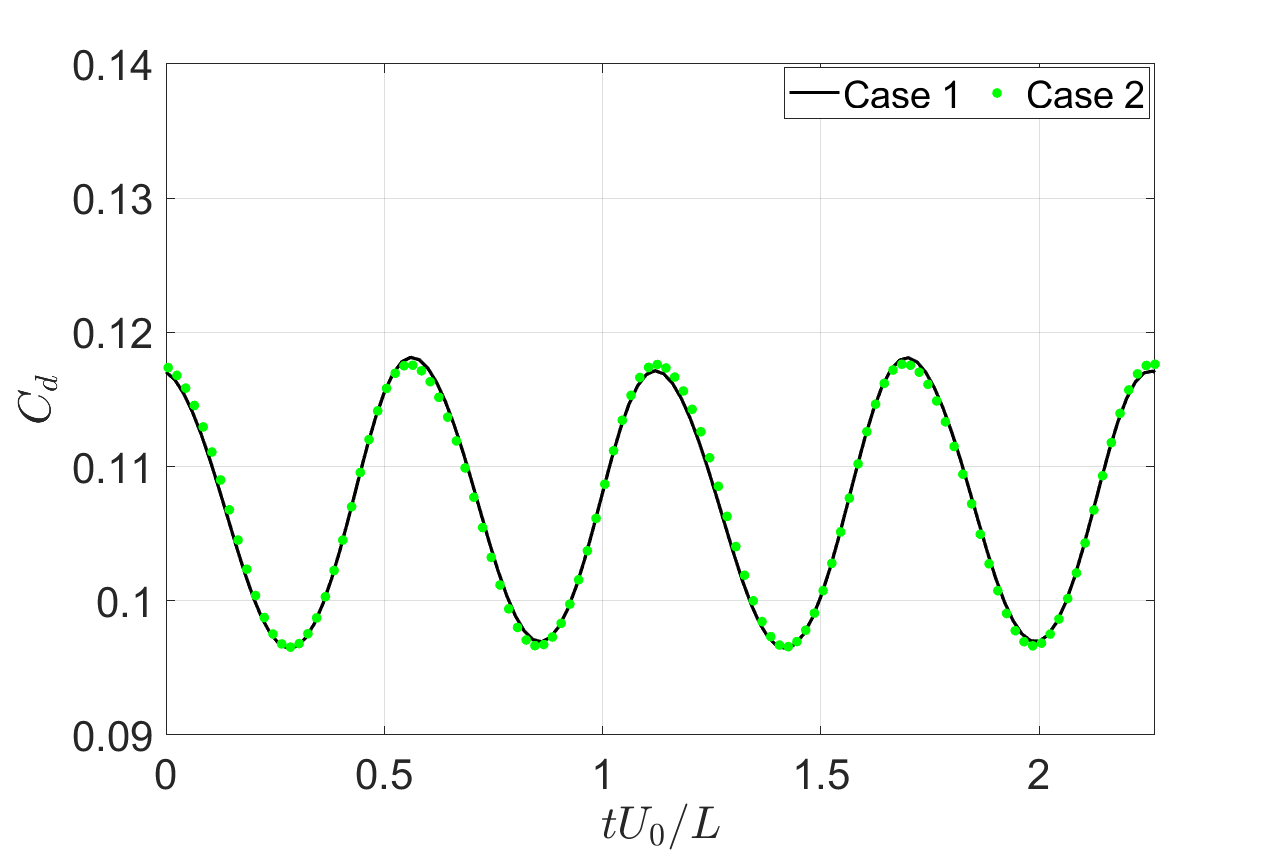}
		\end{minipage}
		\caption{Temporal evolution of the co-efficient of lift (left) and co-efficient of drag (right) over four cycles corresponding to cases 1 and 2.}
		\label{cdcl12} 	
\end{figure}

To validate the equivalent flexural rigidity, we have compared the tip displacement time history for case 1 and case 2. The overlapping plots in Fig.~\ref{tipdisp12} show that the plate tip position is the same at corresponding time instants for both cases. Therefore, the flapping amplitude and frequency response are identical for both cases. Furthermore, a substantial overlap is also observed for the time histories of the lift and drag coefficients of the plate in cases 1 and 2, as seen in Fig.~\ref{cdcl12}. The force profile for the plate in both cases is analogous, resulting in similar flapping dynamics. Table~\ref{results} is a compilation of the maximum, root mean square values of tip displacement along with the mean, maximum, and root mean square values of the lift and drag coefficients. The difference between corresponding values is of the order $10^{-3}$ or less.

\begin{figure}		
		\begin{minipage}[b]{0.49\linewidth}
			\centering
			\includegraphics[trim={2mm 2mm 2mm 2mm},clip,width=0.99\columnwidth]{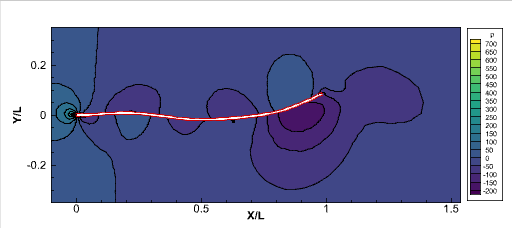}
		\end{minipage}	
		\begin{minipage}[b]{0.49\linewidth}
			\centering
			\includegraphics[trim={2mm 2mm 2mm 2mm},clip,width=0.99\columnwidth]{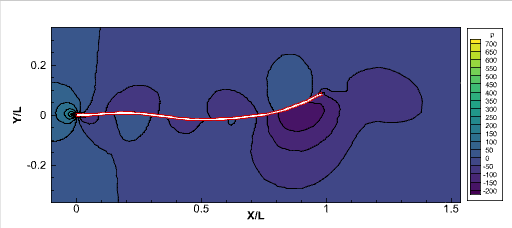}
		\end{minipage}\\
            \begin{minipage}[b]{0.49\linewidth}
			\centering
			\includegraphics[trim={2mm 2mm 0mm 2mm},clip,width=0.99\columnwidth]{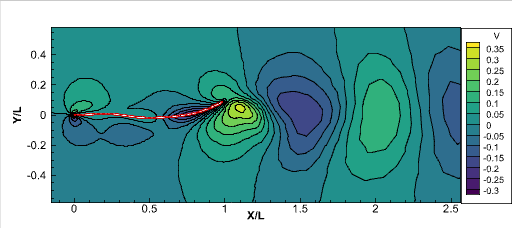}
		\end{minipage}	
		\begin{minipage}[b]{0.49\linewidth}
			\centering
			\includegraphics[trim={2mm 2mm 0mm 2mm},clip,width=0.99\columnwidth]{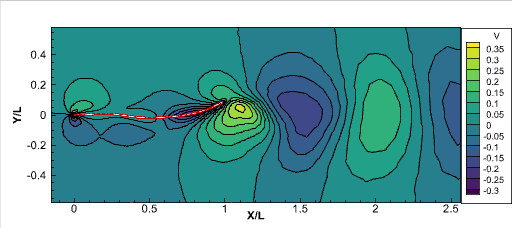}
		\end{minipage}\\
            \begin{minipage}[b]{0.49\linewidth}
			\centering
			\includegraphics[trim={4mm 2mm 2mm 2mm},clip,width=0.99\columnwidth]{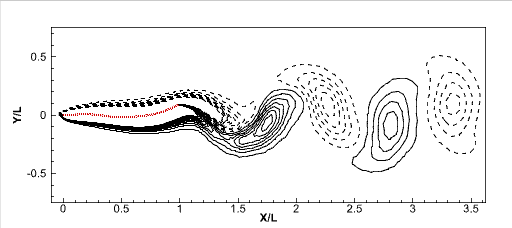}
		\end{minipage}	
		\begin{minipage}[b]{0.49\linewidth}
			\centering
			\includegraphics[trim={4mm 2mm 2mm 2mm},clip,width=0.99\columnwidth]{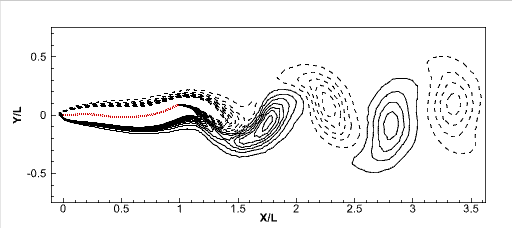}
		\end{minipage}
             \caption{Pressure, velocity and vorticity contours respectively for cases 1 (left) and 2 (right).}
		\label{ff12} 	
\end{figure}

In addition to the similarity of the dynamic behavior of the plate, the flow around the plates also displays strong similitude for cases 1 and 2. Figure~\ref{ff12} shows a side-by-side comparison of the pressure, velocity and vorticity contours for both cases at the same time instant. We can observe the semblance of flow characteristics at matching time instants. Based on the similarities in both cases, we can conclude that the flapping dynamics of the plate depend on the plate's equivalent flexural rigidity, notwithstanding the values of the flexural rigidity for the individual layers.

\begin{figure}
            \centering
            \begin{minipage}[b]{0.49\linewidth}
			\centering
			\includegraphics[trim={1.5mm 2mm 12mm 0mm},clip,width=0.99\columnwidth]{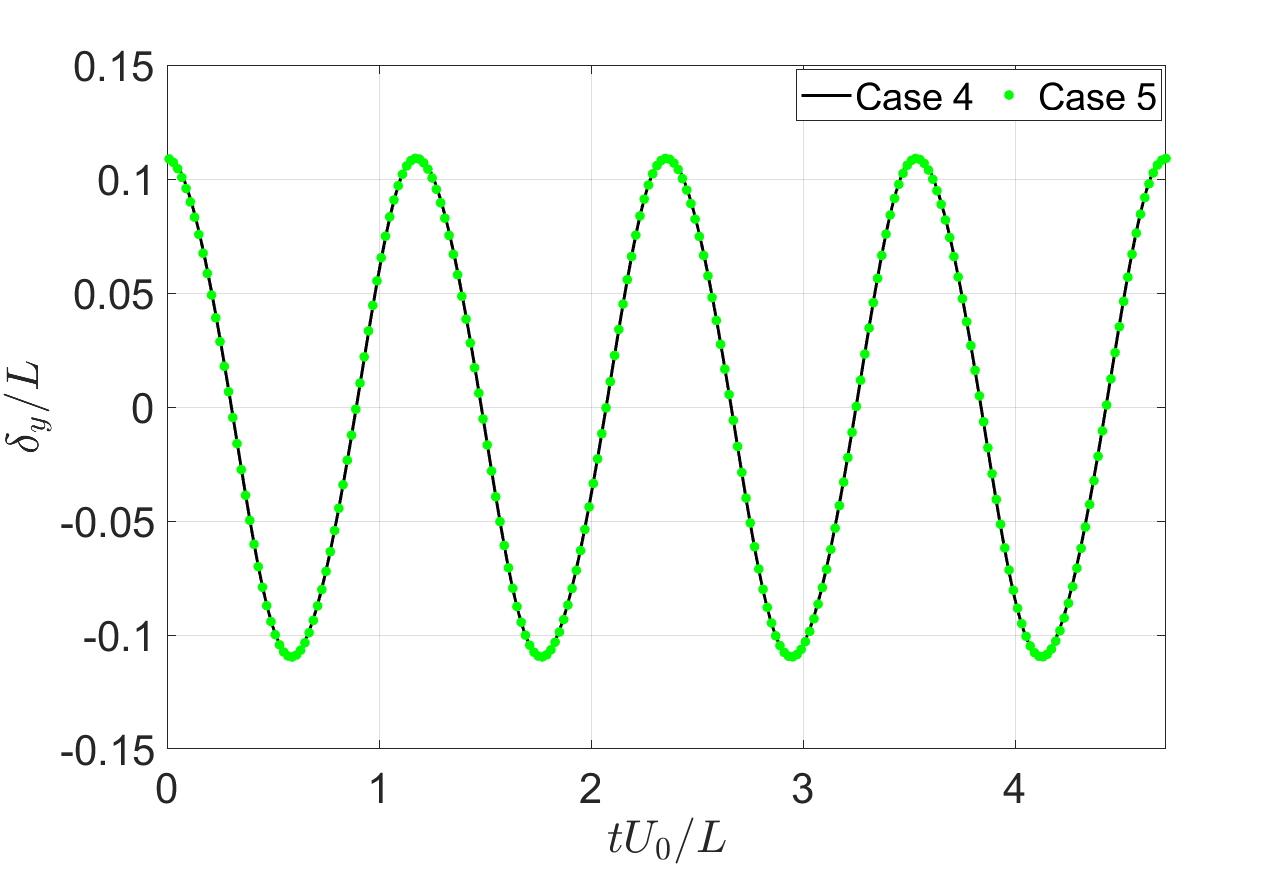}
		\end{minipage}
		\caption{Tip displacement time history for cases 4 and 5, plotted against non-dimensional time.}
		\label{tipdisp45} 	
\end{figure}

\begin{figure}		
            \begin{minipage}[b]{0.49\linewidth}
			\centering
			\includegraphics[trim={1.5mm 2mm 12mm 0mm},clip,width=0.99\columnwidth]{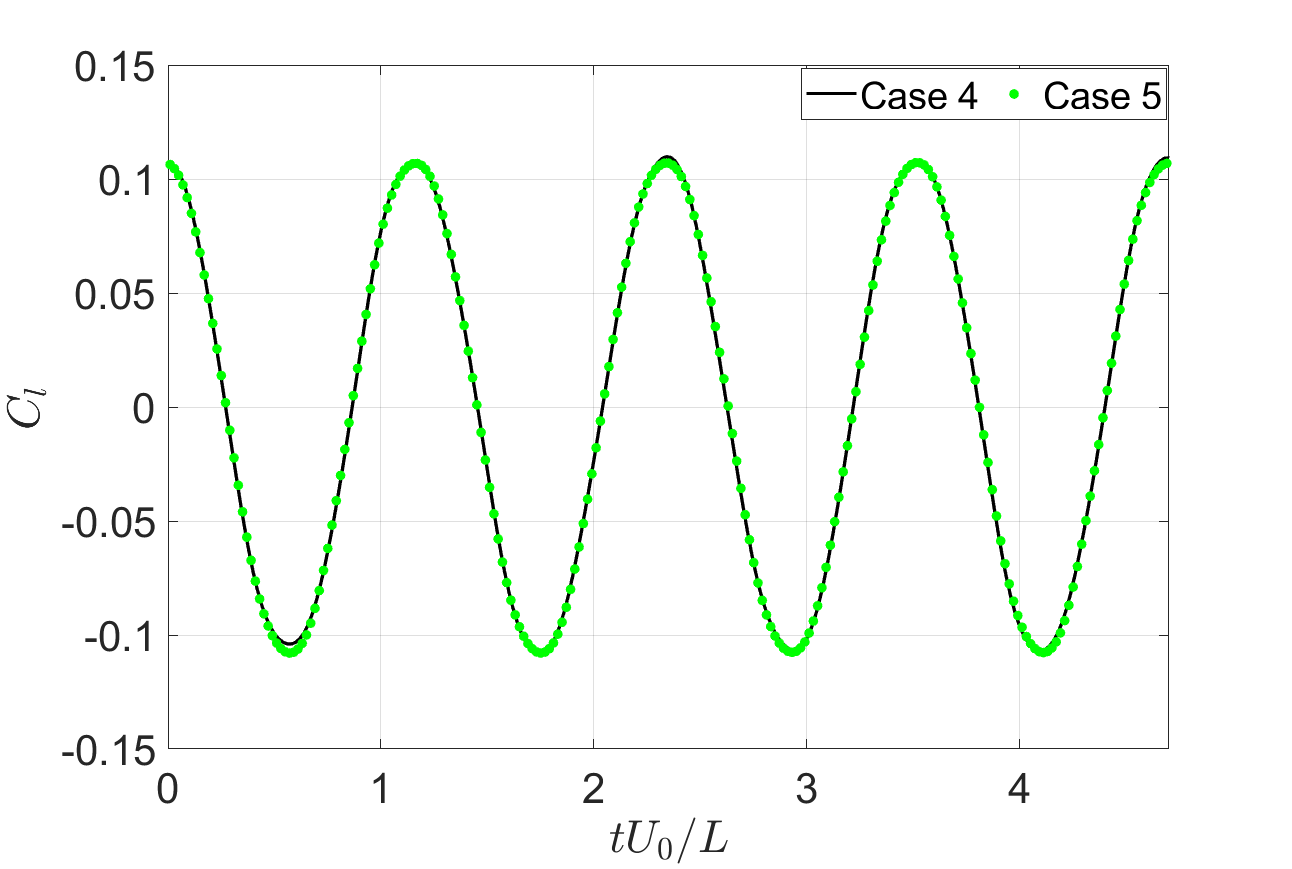}
		\end{minipage}
            \begin{minipage}[b]{0.49\linewidth}
			\centering
			\includegraphics[trim={1.5mm 2mm 12mm 0mm},clip,width=0.99\columnwidth]{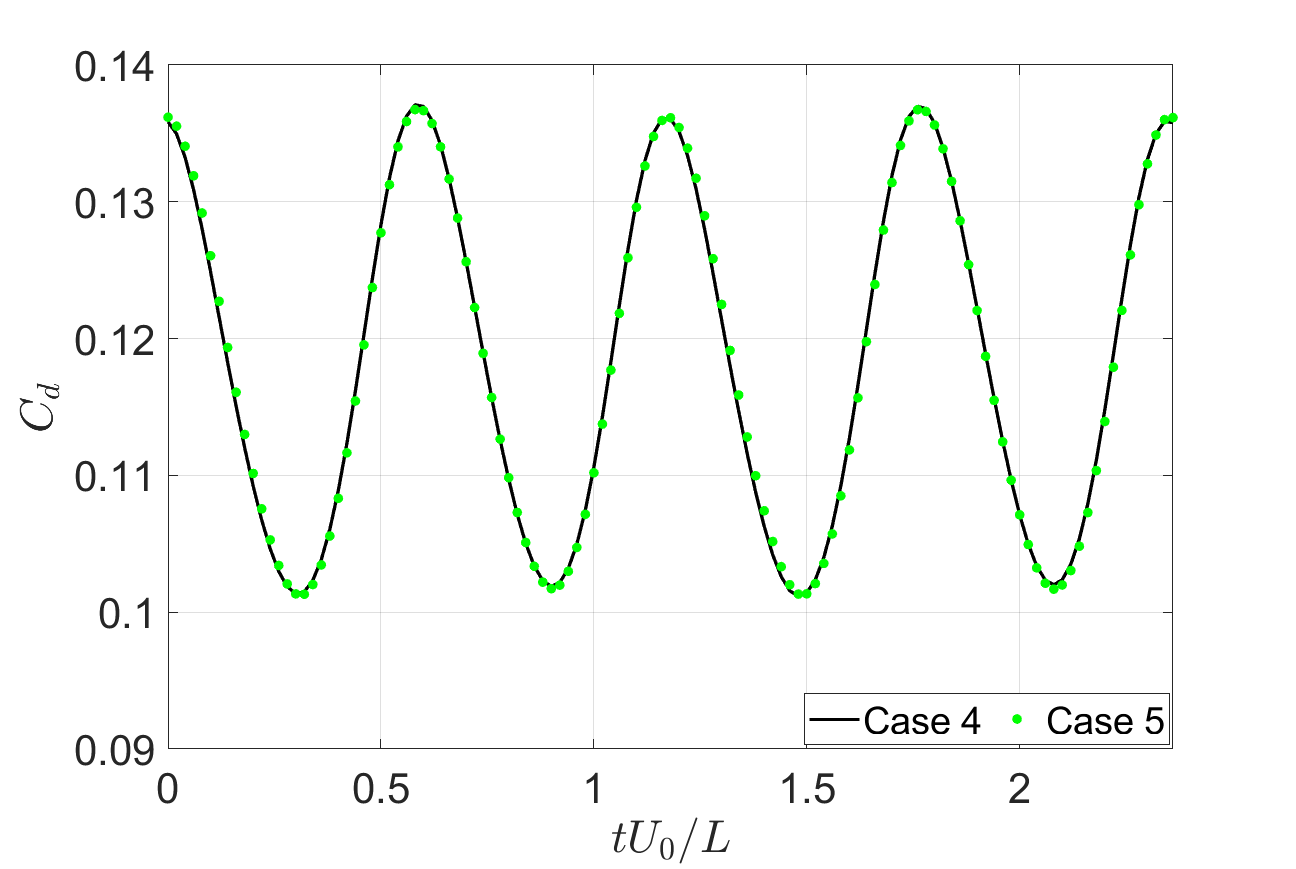}
		\end{minipage}
		\caption{Temporal evolution of the co-efficient of lift (left) and co-efficient of drag (right) over four cycles corresponding to cases 4 and 5.}
		\label{cdcl45} 	
\end{figure}

We have carried out the same exercise for cases 3 and 4. The flow features have been plotted at chronological, equally-spaced time instants over a half-cycle. Details of all numerical results are summarized in the supplementary material. The tip displacement and $C_L$, $C_D$ time histories for cases 4 and 5 are presented in Figs.~\ref{tipdisp45} and ~\ref{cdcl45} respectively. An exact overlap has also been seen in this case. Therefore, the equivalent flexural rigidity formulation holds good even for unequal layer thicknesses. 


\subsection{Validation of equivalent mass ratio}
\begin{figure}
            \centering
            \begin{minipage}[b]{0.49\linewidth}
			\centering
			\includegraphics[trim={1.5mm 2mm 12mm 0mm},clip,width=0.99\columnwidth]{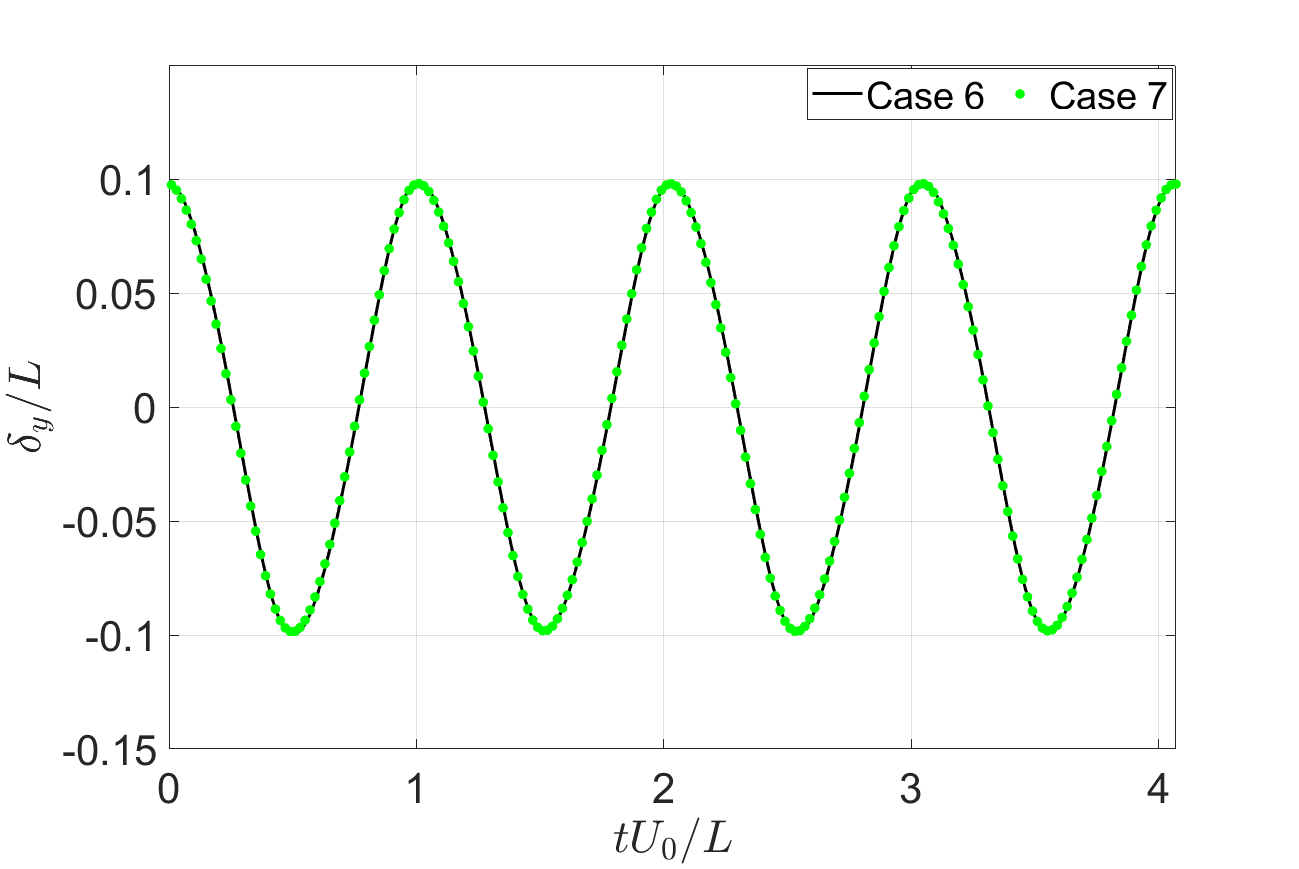}
		\end{minipage}
		\caption{Tip displacement time history for cases 6 and 7, plotted against non-dimensional time.}
		\label{tipdisp67} 	
\end{figure}
\begin{figure}		
            \begin{minipage}[b]{0.49\linewidth}
			\centering
			\includegraphics[trim={1.5mm 2mm 12mm 0mm},clip,width=0.99\columnwidth]{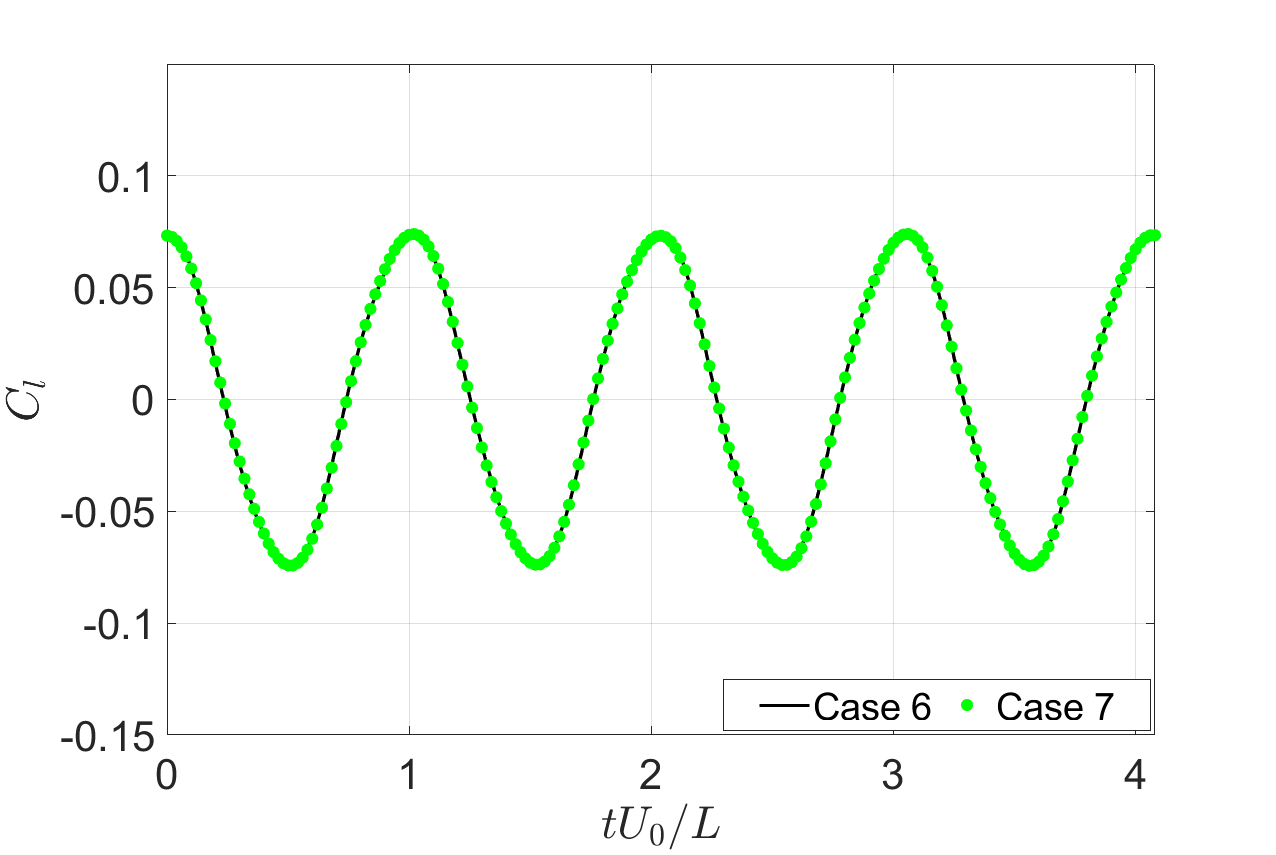}
		\end{minipage}
            \begin{minipage}[b]{0.49\linewidth}
			\centering
			\includegraphics[trim={1.5mm 2mm 12mm 0mm},clip,width=0.99\columnwidth]{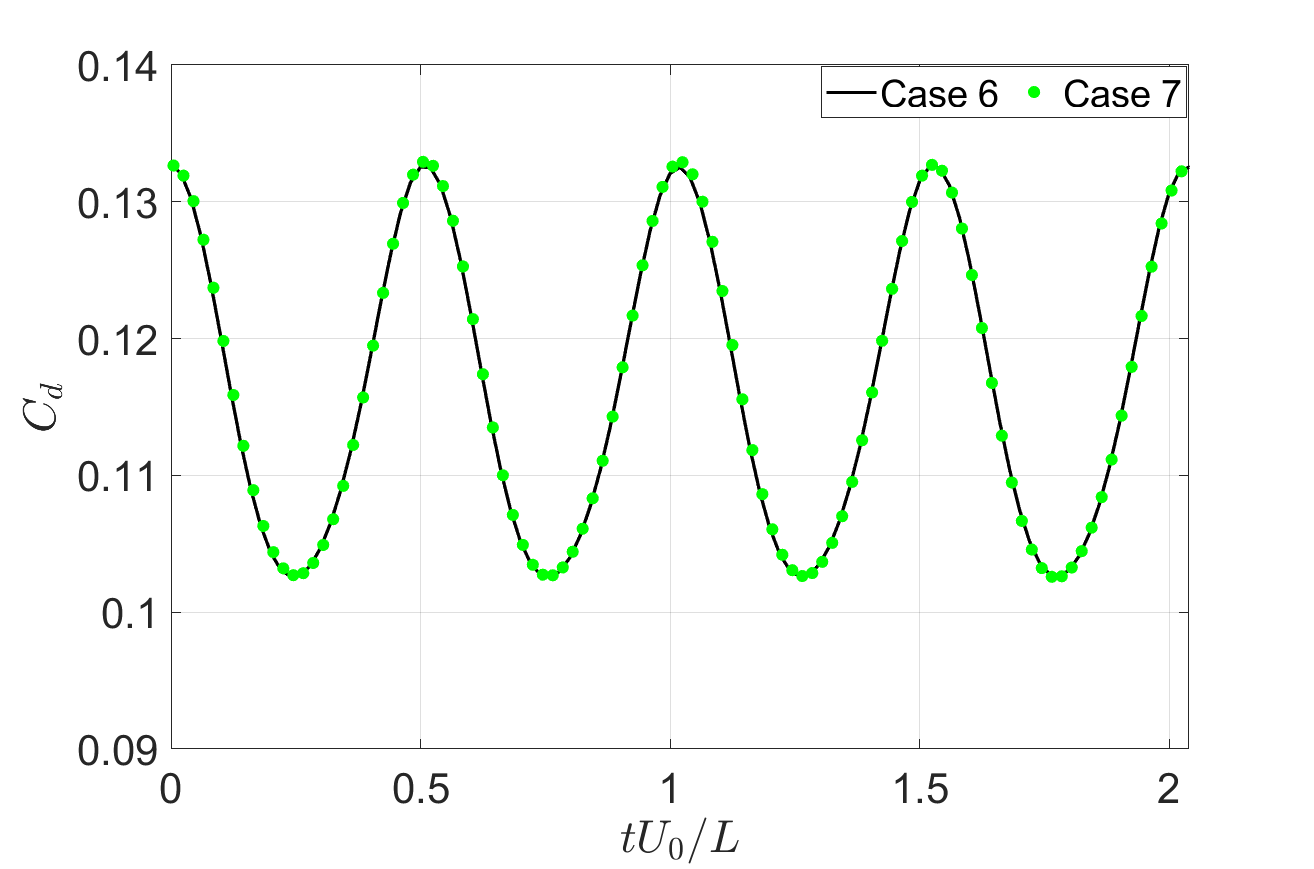}
		\end{minipage}
		\caption{Temporal evolution of the co-efficient of lift (left) and co-efficient of drag (right) over four cycles corresponding to cases 6 and 7.}
		\label{cdcl67} 	
\end{figure}
\begin{figure}		
		\begin{minipage}[b]{0.49\linewidth}
			\centering
			\includegraphics[trim={2mm 2mm 2mm 2mm},clip,width=0.99\columnwidth]{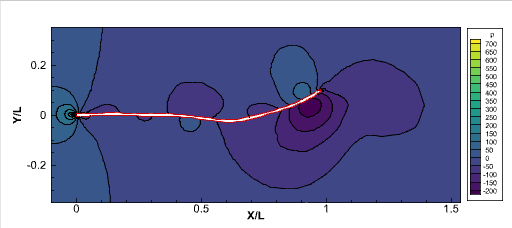}
		\end{minipage}	
		\begin{minipage}[b]{0.49\linewidth}
			\centering
			\includegraphics[trim={2mm 2mm 2mm 2mm},clip,width=0.99\columnwidth]{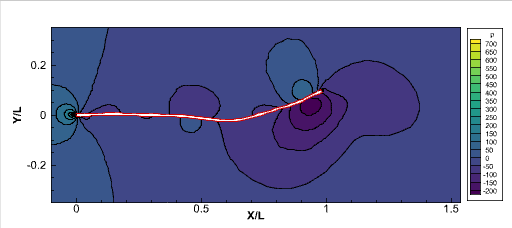}
		\end{minipage}\\
            \begin{minipage}[b]{0.49\linewidth}
			\centering
			\includegraphics[trim={2mm 2mm 0mm 2mm},clip,width=0.99\columnwidth]{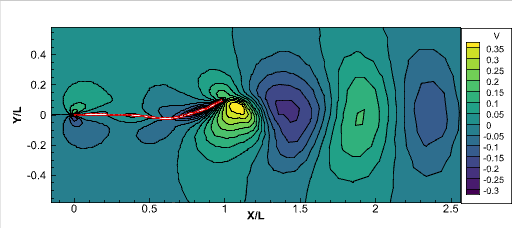}
		\end{minipage}	
		\begin{minipage}[b]{0.49\linewidth}
			\centering
			\includegraphics[trim={2mm 2mm 0mm 2mm},clip,width=0.99\columnwidth]{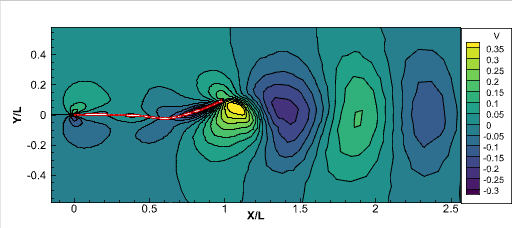}
		\end{minipage}\\
            \begin{minipage}[b]{0.49\linewidth}
			\centering
			\includegraphics[trim={4mm 2mm 2mm 2mm},clip,width=0.99\columnwidth]{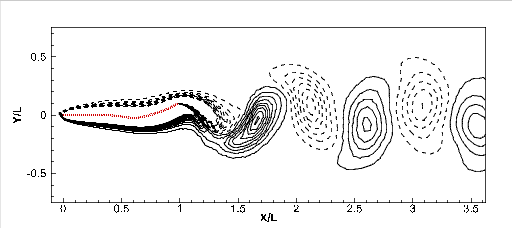}
		\end{minipage}	
		\begin{minipage}[b]{0.49\linewidth}
			\centering
			\includegraphics[trim={4mm 2mm 2mm 2mm},clip,width=0.99\columnwidth]{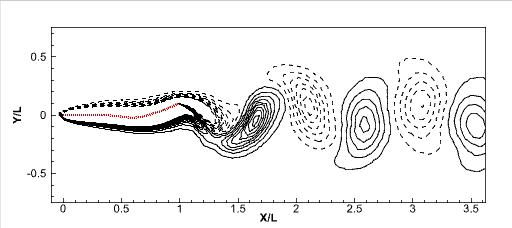}
		\end{minipage}
             \caption{Pressure, velocity and vorticity contours respectively for cases 6 (left) and 7 (right).}
		\label{ff67} 	
\end{figure}
Following the same procedure to validate the equivalent mass ratio, we present the flapping dynamics and flow characteristics for cases 6 and 7. Figure~\ref{tipdisp67} shows the tip displacement time history for both cases. The plots exhibit an exact overlap leading to parity of amplitude and frequency of flapping. The agreement in the values of $C_L$ and $C_D$ of the plate in cases 6 and 7 for identical time instants is depicted in Fig.~\ref{cdcl67}. A quantitative summary for these plots is available in Table~\ref{results}.

\begin{figure}
            \centering
            \begin{minipage}[b]{0.49\linewidth}
			\centering
			\includegraphics[trim={1.5mm 2mm 12mm 0mm},clip,width=0.99\columnwidth]{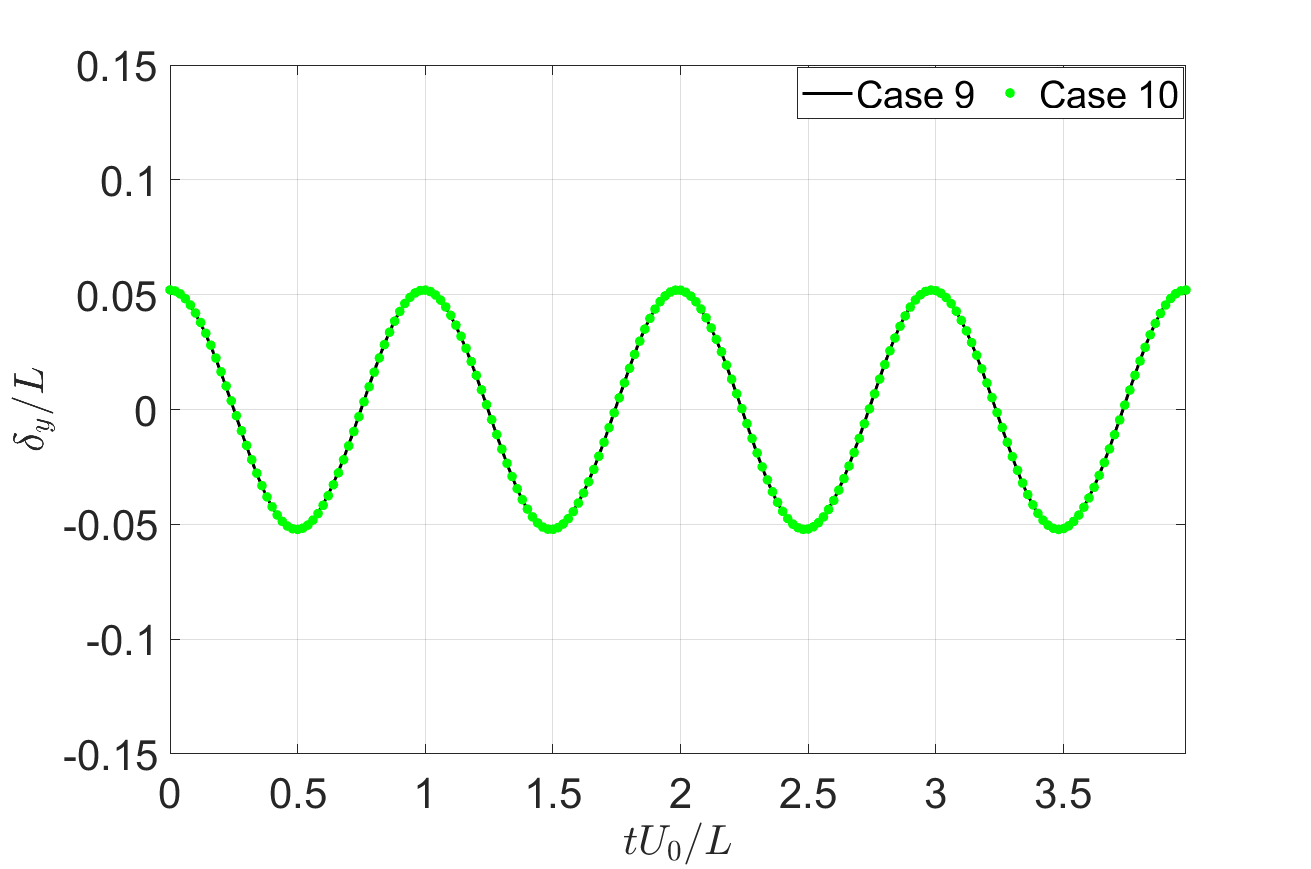}
		\end{minipage}
		\caption{Tip displacement time history for cases 9 and 10, plotted against non-dimensional time.}
		\label{tipdisp910} 	
\end{figure}

\begin{figure}		
            \begin{minipage}[b]{0.49\linewidth}
			\centering
			\includegraphics[trim={1.5mm 2mm 12mm 0mm},clip,width=0.99\columnwidth]{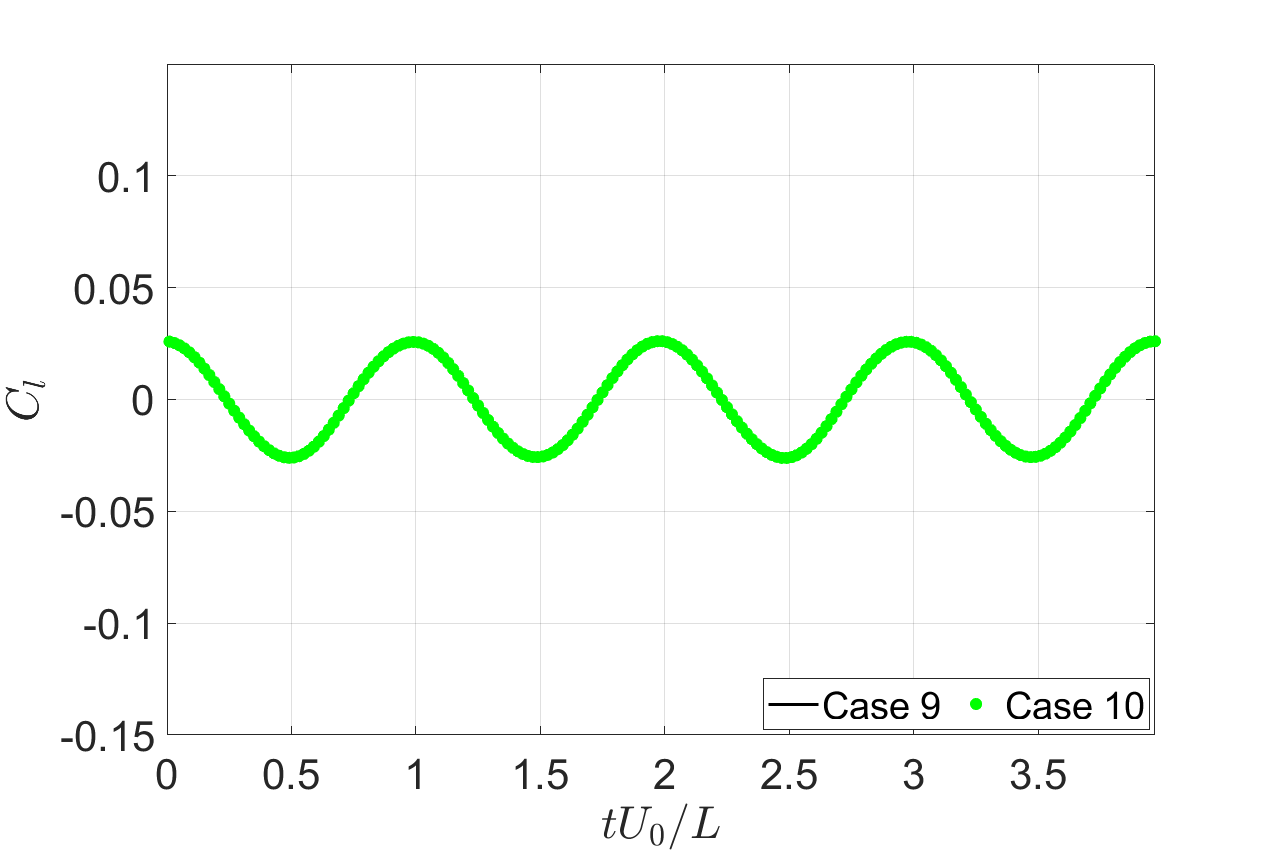}
		\end{minipage}
            \begin{minipage}[b]{0.49\linewidth}
			\centering
			\includegraphics[trim={1.5mm 2mm 12mm 0mm},clip,width=0.99\columnwidth]{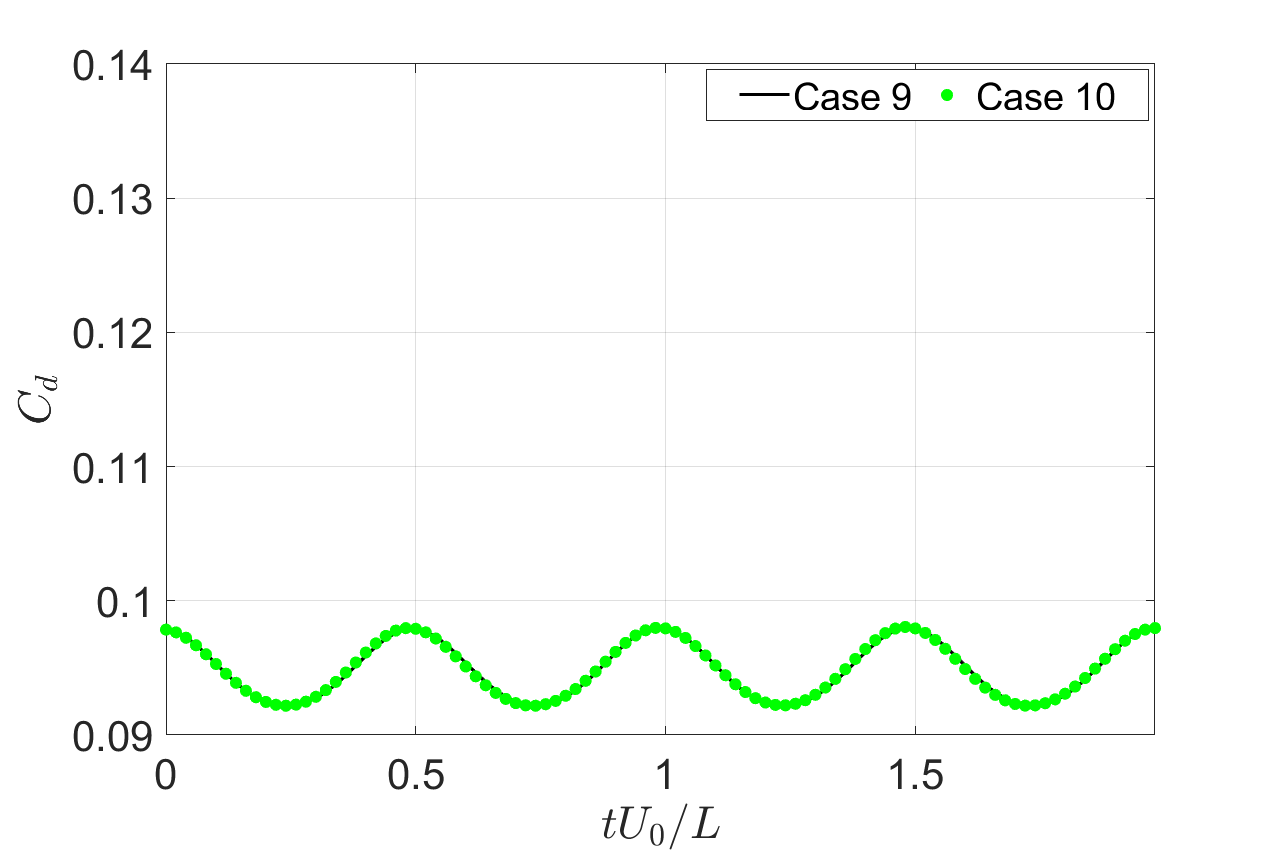}
		\end{minipage}
		\caption{Temporal evolution of the co-efficient of lift (left) and co-efficient of drag (right) over four cycles corresponding to cases 9 and 10.}
		\label{cdcl910} 	
\end{figure}

The pressure, velocity and vorticity contours for cases 6 and 7 are presented in Fig.~\ref{ff67}. Similar flow features are observed for corresponding plate configurations. Cases 8 and 9 also display similarities in the plate's motion and the surrounding fluid's behavior with respect to time. For case 10, we have considered an arbitrary ratio for the layer thicknesses that is not equal to 1. Figures~\ref{tipdisp910} and \ref{cdcl910} present the tip displacement, $C_L$ and $C_D$ time histories respectively, for cases 9 and 10. Here as well, an exact overlap has been observed. The sufficiently strong agreement of the results substantiates the similarity between the cases with identical non-dimensional parameters; thus, the newly proposed non-dimensional parameters effectively predict the flapping dynamics of multilayered plates.

\begin{table}
\caption{Comparison of the statistical data for tip displacement, coefficient of drag and coefficient of lift of the plate obtained from numerical simulations}
\begin{tabular}{c|c|c|c|c|c|c|c}
				Case & $y_{max}$ & $y_{rms}$ & $C_{d}^{\mathrm{mean}}$ & $C_{d}^{\mathrm{max}}$ & $C_{d}^{\mathrm{rms}}$ & $C_{l}^{\mathrm{max}}$ & $C_{l}^{\mathrm{rms}}$\tabularnewline
				\hline 
                1 & 0.08529 & 0.05967 & 0.10709 & 0.11814 & 0.00744 & 0.08095 & 0.05890\tabularnewline
				2 & 0.08524 & 0.06008 & 0.10717 & 0.11770 & 0.00749 & 0.08019 & 0.05939\tabularnewline
                3 & 0.10943 & 0.07692 & 0.11827 & 0.13662 & 0.01234 & 0.10736 & 0.07763\tabularnewline
				4 & 0.10982 & 0.07685 & 0.11837 & 0.13734 & 0.01233 & 0.10969 & 0.07772\tabularnewline
				5 & 0.10913 & 0.07671 & 0.11822 & 0.13684 & 0.01230 & 0.10730 & 0.07757\tabularnewline
				6 & 0.09881 & 0.06864 & 0.11634 & 0.13261 & 0.01063 & 0.07376 & 0.05230\tabularnewline
				7 & 0.09821 & 0.06855 & 0.11634 & 0.13263 & 0.01062 & 0.07359 & 0.05231\tabularnewline
				8 & 0.05204 & 0.03659 & 0.09491 & 0.09803 & 0.00205 & 0.02602 & 0.01836\tabularnewline
				9 & 0.05204 & 0.03604 & 0.09490 & 0.09804 & 0.00204 & 0.02616 & 0.01838\tabularnewline
				10 & 0.05207 & 0.03645 & 0.09489 & 0.09806 & 0.00204 & 0.02611 & 0.01841\tabularnewline
			\end{tabular}
            \label{results}
\end{table}

\section{Conclusion}
\label{conclusion}
In this work, the definition of non-dimensional parameters that describe the flapping dynamics of an isotropic flexible plate placed in a uniform flow has been generalized for the multilayered plate. The equivalent flexural rigidity and the equivalent mass ratio are analytically derived to accommodate the shifted neutral axis. The proposed parameters are validated using numerical simulations to investigate the flapping dynamics of a two-layered plate in uniform flow. The observations prove the ability of the non-dimensional parameters to describe the cross-stream flapping and fluid flow dynamics of multilayered plates with great accuracy. Consequently, the suggested non-dimensional framework can be used to extend the existing experimental, computational, and analytical research on single-layered plates to predict the flapping dynamics of multilayered plates.

\section{CRediT authorship contribution statement}
\textbf{Esha Jain:} Validation, Formal Analysis, Investigation, Writing - Original Draft, Visualization.
\textbf{Aditya Karthik Saravanakumar:} Software, Validation.
\textbf{V. Joshi:} Writing - Review \& Editing.
\textbf{P. S. Gurugubelli:} Conceptualization, Methodology, Software, Supervision.

\section{Declaration of Competing Interest}
The authors declare that they have no known competing financial interests or personal relationships that could have appeared to influence the work reported in this paper.

\section{Acknowledgments}
The corresponding author would like to acknowledge the financial support from the Science and Engineering Research Board's Start-up Research Grant (SERB-SRG) with sanction order number SRG/2019/001249 and BITS Pilani's OPERA award.


\bibliographystyle{elsarticle-num-names} 
\bibliography{refs}

\end{document}